\documentclass[
    ,final            
  ]
  {aipproc}

\layoutstyle{6x9}
\usepackage{amsmath}

\begin{document}

\title{Dynamics of magnetized spherical accretion flows}


\author{Roman Shcherbakov}{
  address={Harvard University, Astronomy Department, 60, Garden st, Cambridge, USA}}

\begin{abstract}
Transonic accretion flow with self-consistent treatment of the
magnetic field is presented. My website
\url{http://www.cfa.harvard.edu/~rshcherb/}.
\end{abstract}

\maketitle

\section{Introduction}
The averaged quantities can be obtained in two different ways in
magnetohydrodynamics. The first way is to solve 3D MHD equations
and then average the results. The second way is to solve some
system of equations on averages. Combination of numerical
simulations and averaged theory brings phenomenology that can
describe observations or experimental data.

The problem of spherically symmetric accretion takes its origin
from Bondi's work \citep{bondi}. He presented idealized
hydrodynamic solution with accretion rate $\dot{M}_B.$ However,
magnetic field $\vec{B}$ always exists in the real systems. Even
small seed $\vec{B}$ amplifies in spherical infall and becomes
dynamically important \citep{schwa}.

Magnetic field inhibits accretion \citep{schwa}. None of many
theories has reasonably calculated the magnetic field evolution
and how it influences dynamics. These theories have some common
pitfalls. First of all, the direction of magnetic field is usually
defined. Secondly, the magnetic field strength is prescribed by
thermal equipartition assumption. In third, dynamical effect of
magnetic field is calculated with conventional magnetic energy and
pressure. All these inaccuracies can be eliminated.

In Section 2\ref{section_method} I develop a model that abandons
equipartition prescription, calculates the magnetic field
direction and strength and employs the correct equations of
magnetized fluid dynamics. In Section 3\ref{results} I show this
accretion pattern to be in qualitative agreement with Sgr A*
spectrum models. I discuss my assumptions in Section 4
\ref{discussion}.

\section{Analytical method}\label{section_method}
 Reasonable turbulence evolution model is the key difference of my
 method. I build an averaged turbulence theory that corresponds to
numerical simulations. I start with the model of isotropic
turbulence that is consistent with simulations of collisional MHD
in three regimes. Those regimes are decaying hydrodynamic
turbulence, decaying MHD turbulence and dynamo action. I introduce
effective isotropization of magnetic field in 3D model.
Isotropization is taken to have a timescale of the order of
dissipation timescale that is a fraction $\gamma\sim1$ of the
Alfven wave crossing time $\tau_{\rm diss}=\gamma r/v_A.$

Common misconception exists about the dynamical influence of
magnetic field. Neither magnetic energy nor magnetic pressure can
represent $\vec{B}$ in dynamics. Correct averaged Euler and energy
equations were derived in \citep{scharlemann} for radial magnetic
field. Magnetic force $\vec{F}_M=[\vec{j}\times\vec{B}]$ can be
averaged over the solid angle with proper combination of
$\vec{\nabla}\cdot\vec{B}=0.$ I extend the derivation to random
magnetic field without preferred direction. Dynamical effect of
magnetic helicity \citep{biskamp03} is also investigated. I
neglect radiative and mechanical transport processes.

The derived set of equations requires some modifications and
boundary conditions to be applicable to the real astrophysical
systems. I add external energy input to turbulence to balance
dissipative processes in the outer flow. The outer turbulence is
taken to be isotropic and has magnetization $\sigma\sim1.$
Transonic smooth solution is chosen as possessing the highest
accretion rate as in \citep{bondi}.

\begin{figure}\label{fig1}
  \includegraphics[height=.5\textheight]{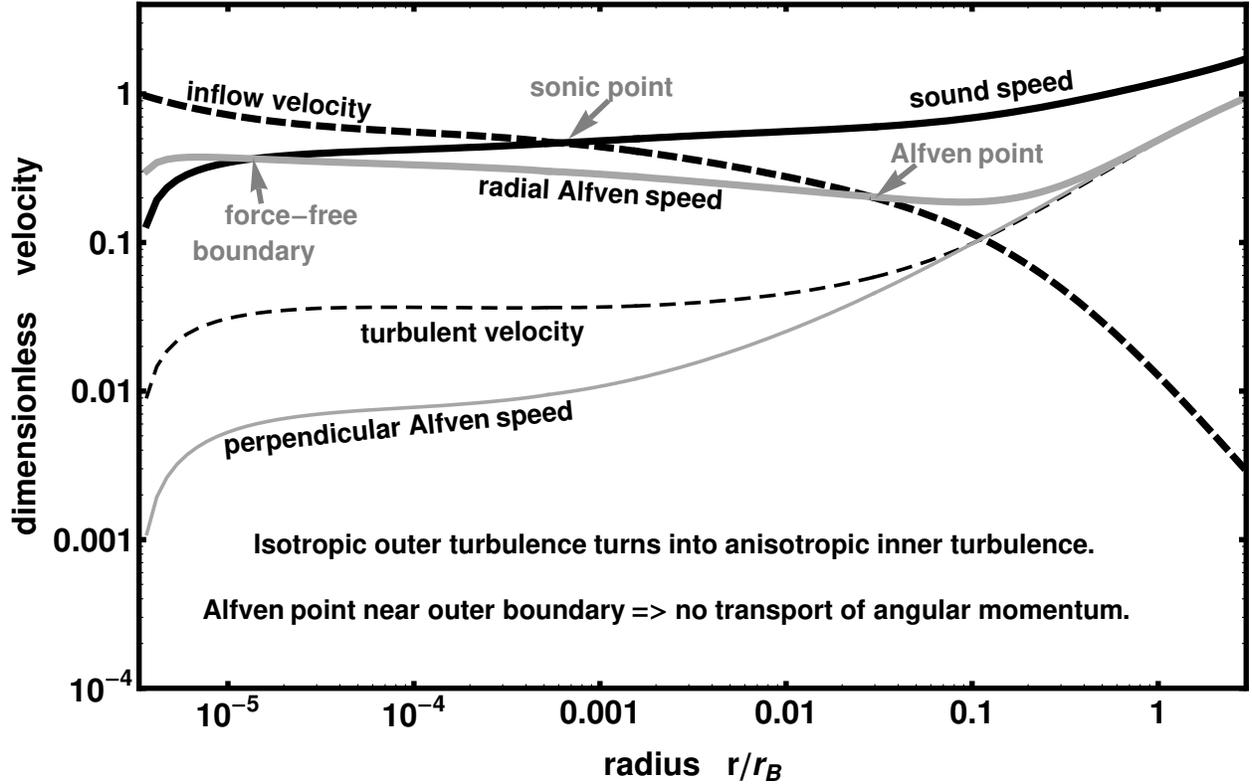}
  \caption{Normalized to Keplerian speed characteristic velocities of magnetized flow. Horizontal lines correspond to self-similar solution $v\sim r^{-1/2}.$}
\end{figure}

\section{Results \& Application to Sgr A*}\label{results}

\begin{figure}\label{fig2}
  \includegraphics[height=.5\textheight]{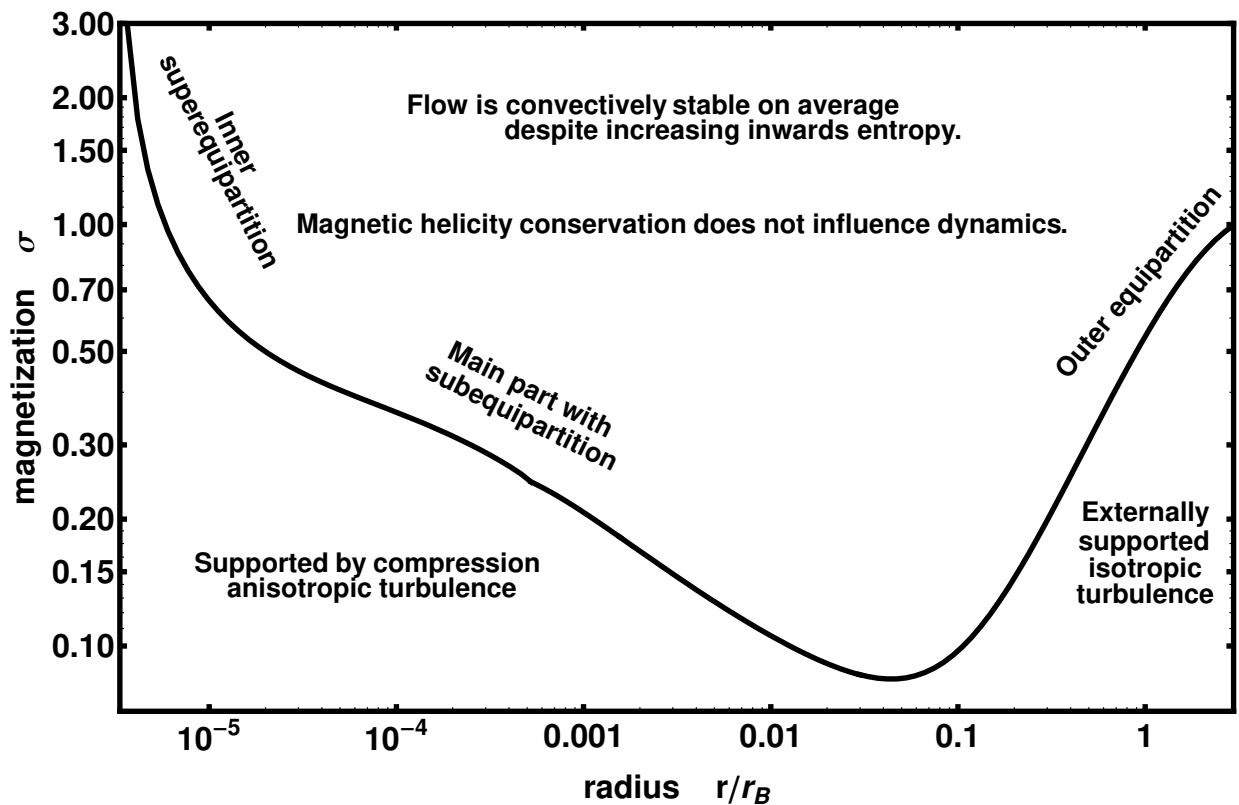}
  \caption{Plot of magnetization $\sigma=(E_M+E_K)/E_{Th}$ with radius.}
\end{figure}
The results of my calculations confirm some known facts about
spherical magnetized accretion, agree with the results of
numerical simulations and have some previously unidentified
features.

Initially isotropic magnetic field exhibits strong anisotropy with
larger radial field $B_r.$ Perpendicular magnetic field
$B_\perp\ll B_r$ is dynamically unimportant in the inner accretion
region Fig\ref{fig1}. Because magnetic field dissipates, infall
onto the black hole can proceed \citep{schwa}.

Turbulence is supported by external driving in the outer flow
regions, but internal driving due to freezing-in amplification
takes over in the inner flow Fig\ref{fig2}. Magnetization of the
flow increases in the inner region with decreasing radius
consistently with simulations \cite{igumen06}. Density profile
appears to be $\rho\sim r^{-1.25}$ that is different from
traditional ADAF scaling $\rho\sim r^{-1.5}$ \citep{narayan}. Thus
the idea of self-similar behavior is not supported.

Compared to non-magnetized accretion, infall rate is 2-5 times
smaller depending on outer magnetization. In turn, gas density is
2-5 times smaller in the region close to the black hole, where
synchrotron radiation emerges \citep{narayan}. Sgr A* produces
relatively weak synchrotron \citep{narayan}. So, either gas
density $n$ or electron temperature $T_e$ or magnetic field $B$
are small in the inner flow or combination of factors works. Thus
low gas density in magnetized model is in qualitative agreement
with the results of modelling the spectrum.

Flow is convectively stable on average in the model of moving
blobs, where dissipation heat is released homogeneously in volume.
Moving blobs are in radial and perpendicular pressure
equilibriums. They are governed by the same equations as the
medium.

\section{Discussion \& Conclusion}\label{discussion}
The presented accretion study self-consistently treats turbulence
in the averaged model. This model introduces many weak assumptions
instead of few strong ones.

I take dissipation rate to be that of collisional MHD simulations.
But flow in question is rather in collisionless regime.
Observations of collisionless flares in solar corona
\citep{noglik} gives dissipation rate $20$ times smaller than in
collisional simulations \citep{biskamp03}. However, flares in
solar corona may represent a large-scale reconnection event rather
than developed turbulence. It is unclear which dissipation rate is
more realistic for accretion.

Magnetic field presents another caveat. Magnetic field lines
should close, or $\vec{\nabla}\cdot\vec{B}=0$ should hold. Radial
field is much larger than perpendicular in the inner region.
Therefore, characteristic radial scale of the flow is much larger
than perpendicular. If radial turbulence scale is larger than
radius, freezing-in condition does not hold anymore. Matter can
freely slip along radial field lines into the black hole. If
matter slips already at the sonic point, the accretion rate should
be higher than calculated.

Some other assumptions are more likely to be valid. Diffusion
should be weak because of high Mach number that approaches unity
at large radius. Magnetic helicity was found to play very small
dynamical role. Only when the initial turbulence is highly
helical, magnetic helicity conservation may lead to smaller
accretion rate. Neglect of radiative cooling is justified a
posteriori. Line cooling time is about $20$ times larger that
inflow time from outer boundary.

The study is the extension of basic theory, but realistic
analytical models should include more physics. The work is
underway.
\begin{theacknowledgments}
I thank my advisor Prof. Ramesh Narayan for fruitful discussions.
\end{theacknowledgments}

\bibliographystyle{aipproc}

\end{document}